\documentclass[aps,prl,twocolumn,superscriptaddress,floats,floatfix]{revtex4}
\usepackage{amssymb,amsmath}
\usepackage{graphics}
\usepackage{epstopdf}
\usepackage{color}
\usepackage{subfigure}
\usepackage{epsfig}
\usepackage{rotating}
\usepackage{float}
\usepackage{anyfontsize}
\usepackage{verbatim}
\usepackage{standalone}
\usepackage{hyperref}

\def\NOTE#1{{\textcolor{red}{\bf [#1]}}}  

\begin{document}
\title{ The formation of compact objects at finite temperatures in
a dark-matter-candidate self-gravitating bosonic system}
\author{Akhilesh Kumar Verma}
\email{akhilesh@iisc.ac.in}
\affiliation{Centre for Condensed Matter Theory, Department of Physics, Indian Institute of Science, Bangalore 560012, India}
\author{Rahul Pandit}
\email{rahul@iisc.ac.in}
\altaffiliation[\\ also at~]{Jawaharlal Nehru Centre For Advanced
Scientific Research, Jakkur, Bangalore, India}
\affiliation{Centre for Condensed Matter Theory, Department of Physics, 
Indian Institute of Science, Bangalore 560012, India} 
\author{Marc E. Brachet} 
\email{brachet@physique.ens.fr}
\affiliation{Laboratoire de Physique de 
l'\'{E}cole Normale Sup{\'e}rieure, 
ENS, Universit\'{e} PSL, CNRS, Sorbonne Universit\'{e}
Universit\'{e} de Paris, F-75005 Paris, France}
\date{\today}
\begin{abstract}

We study self-gravitating bosonic systems, candidates for dark-matter halos, by
carrying out a suite of direct numerical simulations (DNSs) designed to
investigate the formation of finite-temperature, compact objects in the
three-dimensional (3D) Fourier-truncated Gross-Pitaevskii-Poisson
equation (GPPE). This truncation allows us to explore the collapse and
fluctuations of compact objects, which form at both zero temperature
and finite temperature. We show that the statistically steady state of
the GPPE, in the large-time limit and for the system sizes we study,
can also be obtained efficiently by tuning the temperature in an
auxiliary stochastic Ginzburg-Landau-Poisson equation (SGLPE). We show
that, over a wide range of model parameters, this system undergoes a
thermally driven first-order transition from a collapsed, compact,
Bose-Einstein condensate (BEC) to a tenuous Bose gas without condensation. By a
suitable choice of initial conditions in the GPPE, we also obtain a
binary condensate that comprises a pair of collapsed objects rotating
around their center of mass.

\end{abstract}
\maketitle

Gravitational effects are important on stellar scales; it might also be
possible to mimic such effects in laboratory Bose-Einstein
condensates~\cite{ODell2000} and thus emulate gravitationally bound, condensed,
assemblies of bosons, which are candidates for dark-matter
halos~\cite{RuffiniBonazzola1969,Ingrosso:1991,Jetzer1992,Hu2000}. Although
many experiments have been carried out to establish the identity of dark
matter, there is still no unambiguous dark-matter candidate. For many years the
front runners have been weakly interacting massive particles (WIMPs); but their
existence has not been established convincingly (see, e.g.,
~\cite{abdallah2018search,kachulis2018search} and references therein), so
investigations of other candidates, e.g., axions, boson stars, black holes, and
superfluids, have experienced a
renaissance~\cite{apslink,lawson2019tunable,Chavanis1,Chavanis2,Chavanis3,Chavanis4,Chavanis5}.

The Gross-Pitaevskii-Poisson equation (GPPE), for a self-gravitating assembly
of weakly interacting bosons, is the natural theoretical model for
dark-matter-candidate bosonic assemblies, both in laboratory and astrophysical
settings. We address finite-temperature ($T > 0$) effects in such condensation,
a problem of central importance in this challenging field.  Several ways have
been suggested for including $T > 0$ effects in the Gross-Pitaevskii (GP)
model~\cite{PhDTutorial} without gravity; one important way uses the
Fourier-truncated GP model, in which this truncation generates a
classical-field model~\cite{PhDTutorial,Berloff14,Krstulovic11,vmrnjp13}.  We
generalize these studies by using the Fourier-truncated GPPE to study $T > 0$
effects during the gravitational collapse of a system of self-gravitating
bosons. \textit{In addition}, we define an algorithm that directly reconstructs
the thermalized state of such a system  by using an auxiliary stochastic
Ginzburg-Landau-Poisson equation (SGLPE).

We obtain several interesting results by pseudospectral direct numerical
simulations (DNSs) of the truncated GPPE and the SGLPE in three dimensions
(3D). We follow the spatiotemporal evolution of different initial conditions
for the density of bosons. If we start from a very-nearly uniform density in
the truncated GPPE, this system undergoes gravitational collapse and
thermalizes to a Bose-Einstein condensate (BEC), at low $T$. Our SGLPE study
yields a hitherto unanticipated, thermally driven, first-order phase transition
from a collapsed, compact, BEC to a tenuous Bose gas without condensation, for
a wide variety of parameters in the GPPE. Finally, by a suitable choice of
initial conditions in the GPPE, we obtain a binary condensate that comprises a
pair of collapsed objects rotating around their center of mass.



\begin{figure*}[ht]
\centering	
\resizebox{\linewidth}{!}{
\includegraphics[scale=0.5]{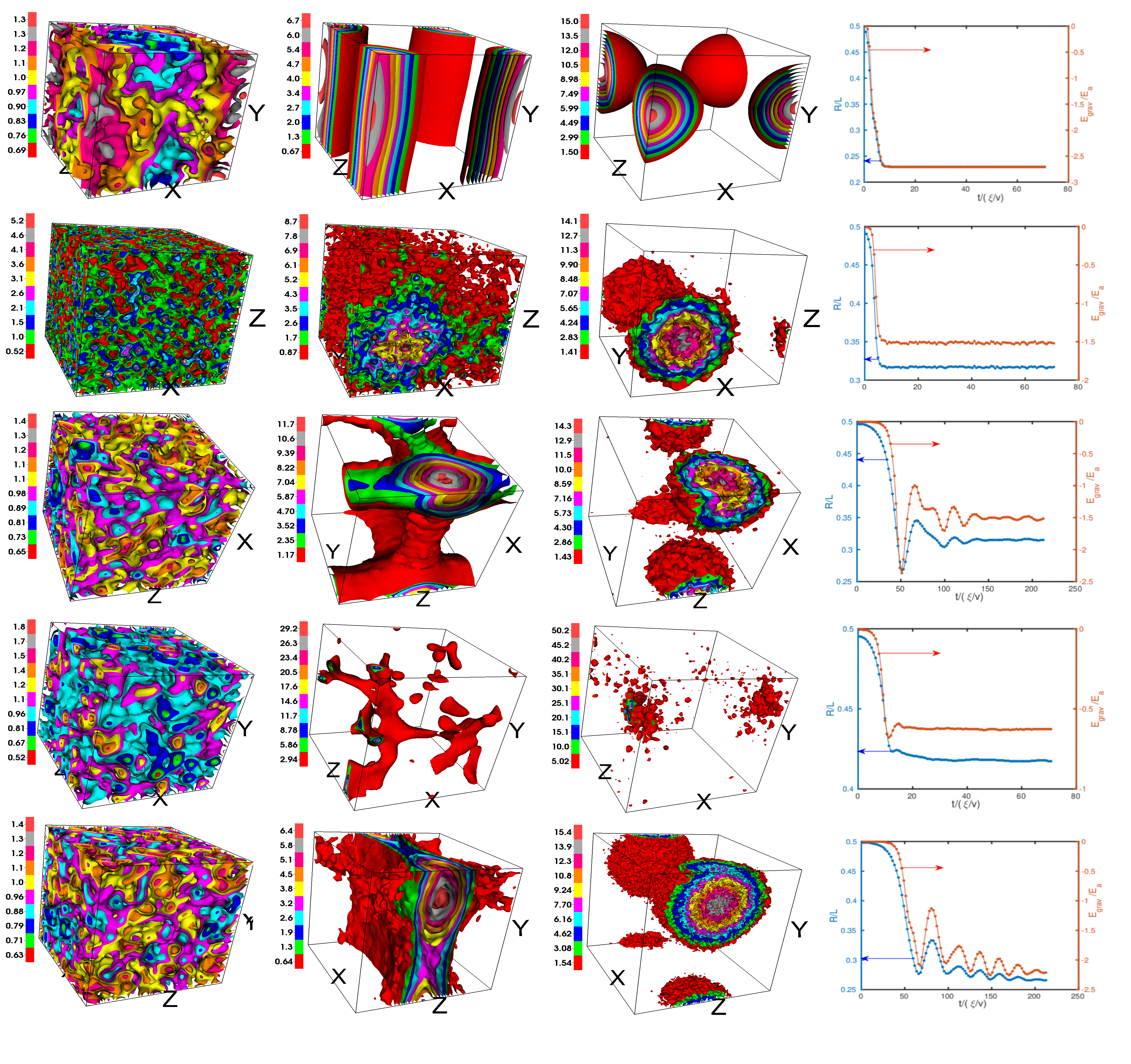}}





\caption{ \small Columns $(1)-(3)$: Ten-level contour plots of
${\left| \psi({\bf x},t) \right|}^2$, at representative times: SGLPE ($T=0$)
(top row, run $R1$); SGLPE (second row, run
$R2$); GPPE (third row, run $R3$); GPPE (fourth row, run $R4$); and 
$256^3$ GPPE (fifth row, run $R5$) [the videos V1-V5 
(Supplemental Material~\cite{supplement}) show, respectively, the complete 
spatiotemporal evolution for these cases]. 
Column$(4)$: Plots of the scaled radius of gyration $\frac{R}{L} =\frac{1}{L}
\sqrt{\frac{\int_V \rho(r) r^2 d{\bf r}}{\int_V \rho(r)d{\bf r}}}$,
(blue) and the scaled gravitational energy $E_{\rm grav}/E_a$
(red ) versus the scaled time $t/(\xi/v)$ for the
different runs, where $E_{\rm a} = 2^5 \pi^4 (G/g^3)^{1/2}$.}\label{fig:big}
\end{figure*}

A self-gravitating BEC is described by a complex wave function $\psi({\bf
x},t)$; for weakly interacting bosons, the spatiotemporal evolution of
$\psi({\bf x},t)$ is governed by the GPPE: 
\begin{eqnarray} 	
i \hbar \partial_t \psi  &=& -\frac{\hbar^2}{2m} \nabla^2 \psi  +\left[G \Phi +
g |\psi|^2\right] \psi, \nonumber \\
\nabla^2 \Phi &=& |\psi|^2  - < |\psi|^2> \label{eq:GPPE},
\end{eqnarray} 	
where $m$ is the mass of the bosons,  $n=|\psi|^2$ their number density,
$\Phi$ is the gravitational potential field, $G = 4\pi G_{\rm N}m^2$ ($G_{\rm N}$ is Newton's constant), and $g=4 \pi a
\hbar^2/m$, with $a$ the $s$-wave scattering length.  The subtraction of the
mean density $\langle|\psi|^2\rangle$  can be justified either by taking into
account the cosmological expansion~\cite{Peebles,falco2013} or by defining a
Newtonian cosmological constant~\cite{KIESSLING2003}.  By linearizing
Eq.~\eqref{eq:GPPE} around the constant $|\psi|^2=n_0$, we obtain the
dispersion relation $\omega(k)=\sqrt{-Gn_0/m+k^2 g n_0/m+k^4 (\hbar/2m)^2}$,
which displays a low-$k$ Jeans instability, for wave numbers  $k < k_{\rm
J}=\sqrt{\frac{G}{g}} \left[\left(1+\sqrt{1+\frac{G \hbar^2}{m g^2
n_0}}\right)/2\right]^{-1/2}$.  In the absence of gravity ($G=0$), we identify
the speed of sound $c=\sqrt{\frac{g n_0}{m}}$ and the coherence length $\xi =
\sqrt{\frac{\hbar^2 }{2g{n_0}m}}$. Units relevant to astrophysics are 
discussed in ~\cite{Chavanis5}.

We solve the GPPE~\eqref{eq:GPPE} by using the $3D$ Fourier pseudospectral
method, with the $2/3$-rule for
dealiasing~\cite{Krstulovic11,vmrnjp13,Got-Ors}: We expand the $2\pi$ periodic
wave function as $\psi(x)=\sum_{{\bf k}\in \mathbb{Z}^3} \hat \psi_{\bf k}
\exp(i {\bf k}\cdot{\bf x})$ and then we truncate it spectrally, by setting
$\hat{\psi}_{\bf k}\equiv0$ for $|{\bf k}| > k_{\rm max}$, with $k_{\rm
max}=[{\cal N}/3]$, where $\cal{N}$ is the resolution and $[\cdot]$ denotes the
integer part~\cite{PhDTutorial,Berloff14}. If we introduce the Galerkin
projector $\mathcal{P}_{\rm G}$ [in Fourier space $\mathcal{P}_{\rm G} [
\hat{\psi}_{\bf k}]=\theta(k_{\rm max}-|{\bf k}|)\hat{\psi}_{\bf k}$,
with $\theta(\cdot)$ the Heaviside function], the Fourier-truncated GPPE
becomes 
\begin{equation}
i\hbar\frac{\partial\psi}{\partial t} =\mathcal{P}_{\rm G} [- \frac{\hbar^2}{2m} {\bf \nabla}^2 \psi + \mathcal{P}_{\rm G} [(G {\bf \nabla}^{-2}+g)|\psi|^2]\psi ].
\label{Eq:TGPPEphys}
\end{equation}
Equation~(\ref{Eq:TGPPEphys}) \textit{conserves exactly} the number of
particles $N=\int d^3 x|\psi|^2$ and the energy $E=E_{kq}+E_{int}+E_{G}$, where
$E_{kq}=\frac{\hbar^2}{2m} \int d^3 x |{\bf \nabla} \psi |^2$,
$E_{int}=\frac{g}{2} \int d^3 x [\mathcal{P}_{\rm G}|\psi|^2]^2 $, and $E_{G}=
E_{\rm grav} = \frac{G}{2} \int d^3 x[\mathcal{P}_{\rm G}|\psi|^2] {\bf
\nabla}^{-2}[\mathcal{P}_{\rm G}|\psi|^2]$. If we use the $2/3$-rule for
dealiasing, then the momentum ${\bf P}=\frac{i\hbar}{2m}\int d^3x\left( \psi
{\bf \nabla}\overline{\psi} - \overline{\psi} {\bf \nabla}\psi\right)$, 
where the over-bar denotes complex conjugation, is also 
conserved~\cite{Krstulovic11}. 

This spectral truncation generates a classical-field model that allows us to
study finite-$T$ effects in the GPPE (Refs.~\cite{Berloff14,Krstulovic11,vmrnjp13} for the GP case).  We
show that this spectrally truncated GPPE can describe dynamical effects and,
\textit{at the same time}, yield thermalized states, which we can obtain both
by the thermalization of the long-time GPPE dynamics or directly, by using the
SGLPE
\begin{eqnarray}
\nonumber \hbar\frac{\partial\psi}{\partial t}&=&\mathcal{P}_{\rm G} [\frac{\hbar^2}{2m} {\bf \nabla}^2 \psi + \mu \psi  - \mathcal{P}_{\rm G} [(G {\bf \nabla}^{-2}+g)|\psi|^2]\psi ] \\
&&+\sqrt{\frac{2 \hbar}{\beta}} \mathcal{P}_{\rm G} \left[\zeta({\bf x},t)\right] \label{eq:SGLPE},\hspace{3mm}
\end{eqnarray}
where the zero-mean, Gaussian white noise $\zeta({\bf x},t)$ has the variance
$\langle\zeta({\bf x},t)\zeta^*({\bf x'},t')\rangle=\delta(t-t') \delta({\bf
x}-{\bf x'})$, with $\beta=\frac{1}{k_B T}$ the inverse temperature; we tune
the chemical potential $\mu$ at each time step to conserve the total number of
particles $N$.  The finite-$T$ SGLPE dynamics does not describe any physical
evolution; but it converges more rapidly, than does the GPPE dynamics, toward a
thermalized state (Refs.~\cite{Berloff14,Krstulovic11,vmrnjp13} for the GP
case). Clearly, the SGLPE leads to a state with a given temperature; but the
GPPE yields a state with a given energy. 

Our pseudospectral DNSs of the GPPE \eqref{eq:GPPE}  and SGLPE \eqref{eq:SGLPE}
use a cubical computational domain that is $(2 \pi)^3$ periodic; we normalise
$\psi$ such that $N=(2\pi)^3$, and use units with $\hbar=1$ and $m=1$.  We list
the parameters for different runs in Tables~\ref{tab:dimensionless} and
\ref{tab:runs}. 

\begin{table}
\resizebox{0.8\linewidth}{!}
{
\begin{tabular}{|l|l|l|}
\hline
$M_{\rm a} = \frac{\hbar}{\sqrt{G_{N}ma}}$ & 
$\frac{M}{M_{\rm a}} = \frac{M \sqrt{G g}}{4 \pi \hbar^2}$ \\
\hline
$ R_{\rm a}=\sqrt{\frac{a \hbar^2}{m^3 G_N}}$ &
$\frac{R}{R_{\rm a}} = \frac{R}{\sqrt{g/G}}$ \\
\hline 
$ a_{\rm Q} = \frac{\hbar^2}{G_{N}M^2m} $ &
$\frac{a}{a_{\rm Q}} = \frac{M^2 {G g}}{(4 \pi \hbar)^2} $ \\
\hline
$  R_{\rm Q} = \frac{\hbar^2}{G_{N}m^2M}$ &
$ \frac{R}{R_{\rm Q}} = \frac{M {G R}}{4 \pi \hbar^2}$ \\	
\hline
\end{tabular}}

\caption{ Dimensionless variables: in the second column of the
table we use quantities from the first column; $G_{\rm N}$ denotes 
Newton's constant and $G = 4\pi G_{\rm N}m^2$.}\label{tab:dimensionless}
\end{table}

\begin{table}
\resizebox{1.0\linewidth}{!}
{\begin{tabular}{|l|l|l|l|l|l|}
\hline
{\rm Run} & R1 & R2 & R3 & R4 & R5 \\
\hline
{\rm Type}  & SGLPE ($T=0$) & SGLPE & GPPE & GPPE & GPPE \\
\hline
$\cal{N}$ & $64$ & $64$ & $64$ & $64$ & $256$ \\
\hline
$g$ & $50$  & $50$ & $50$ & $5$ & $50$ \\
\hline
$G$ & $105$ & $105$ & $105$ & $550$ & $105$ \\
\hline
\end{tabular}}
\caption{\small This table shows the representative runs for
which we give plots in Fig.$1$.}\label{tab:runs}
\end{table}

\begin{figure*}[ht]
\resizebox{\linewidth}{!}{
\includegraphics[scale=0.5]{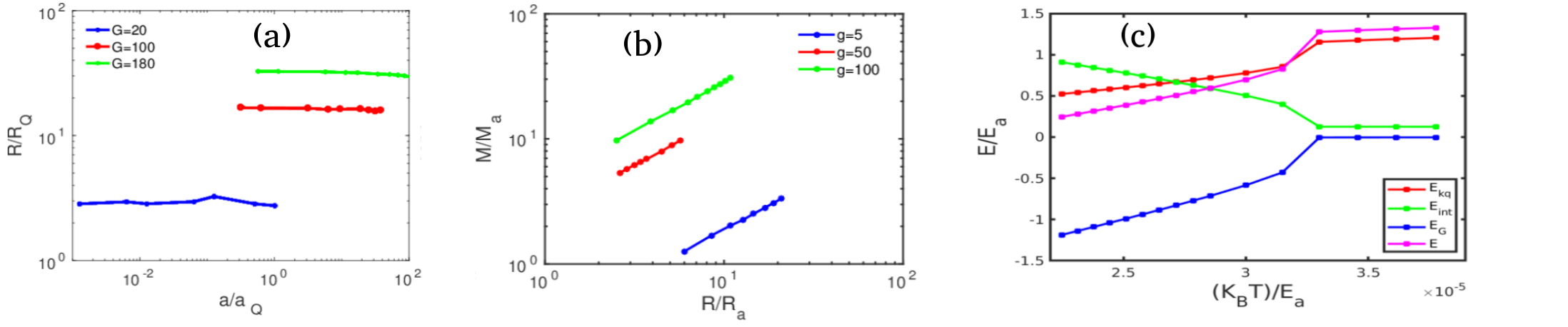}}
\caption{The plots of (a) the scaled radius of gyration $(R/R_{\rm Q})$ versus 
scaled scattering length $(a/a_{\rm Q})$ for the GPPE Runs A1-A30, (b) the scaled mass 
$(M/M_{\rm a})$ versus the scaled radius of gyration $R/R_{\rm a}$ for GPPE Runs
B1-B30 on a log-log scale, (c) the scaled energy components ($E_{\rm kq}/E_{\rm a},
E_{\rm int}/E_{\rm a}, E_{\rm G}/E_{\rm a})$, and the total energy 
$(E/E_{\rm a})$ versus the  scaled temperature $k_{\rm B}T/E_{\rm a}$.}
\label{fig:GPPE}
\end{figure*}

In columns $(1)-(3)$ of Fig.~\ref{fig:big} we show ten-level contour plots of
${\left| \psi({\bf x},t) \right|}^2$ to illustrate, at representative times,
the spatial organization of ${\left| \psi({\bf x},t) \right|}^2$ that we obatin
via the SGLPE ($T=0$) (top row, run $R1$), the SGLPE (second row, run $R2$),
and three GPPE runs (third row, run $R3$; fourth row, run $R4$; fifth row, run
$R5$); the videos V1-V5 (Supplemental Material~\cite{supplement}) show,
respectively, the complete spatiotemporal evolution of  ${\left| \psi({\bf
x},t) \right|}^2$ for these five runs.  In column $(4)$ of Fig.~\ref{fig:big}
we give, for these runs, plots of the scaled radius of gyration $R/L =
\frac{1}{L}\sqrt{\frac{\int_V \rho(r) r^2 d{\bf r}}{\int_V \rho(r)d{\bf r}}}$
(blue curves) and the scaled gravitational energy $E_{\rm grav}/E_a$ (red
curves) versus the scaled time $t/(\xi/v)$, where $E_{\rm a} = 2^5 \pi^4
(G/g^3)^{1/2}$. If we tune $T$ in the SGLPE~\eqref{eq:SGLPE}, it yields a
statistically steady state whose properties (like $R/L$ and $E_{\rm grav}/E_a$)
are close to their counterparts in the thermalized state of the GPPE (e.g., by
comparing rows (2) and (3) in  column(4) of Fig.~\ref{fig:big} we see that the
final state of the SGLPE has nearly the same energy and radius as the
corresponding statistically steady GPPE state).  Furthermore, convergence to
this thermalized state is more rapid in the (canonical) SGLPE than in the
(microcanonical) GPPE.  To validate our DNSs, we have checked explicitly
(Supplemental Material~\cite{supplement}) that, at zero temperature, our
results agree with those of the $T=0$ study of
Refs.~\cite{Chavanis1,Chavanis2}, which yield spherically symmetric ground
states ( row $(1)$ of Fig.~\ref{fig:big}) with radius $R$ and $N$ bosons.
Their ground-state energy can be approximated as $E(R)=\frac{\hbar^2 N}{2 m
R^2}+\frac{g N^2}{2 R^3}-\frac{G N^2}{2 R}$ and the equilibrium radius follows
from $dE/dR|_{R=R_0}=0$, whence $R_0=\frac{R_{\rm Q}}{4\pi}(1+\sqrt{1+48 \pi^2
(\frac{R_{\rm a}}{R_{\rm Q}}})^2$, where $R_{\rm Q}=\frac{\hbar^2} {G_{\rm N}
m^3 N}$ and $R_{\rm a}=\sqrt{\frac{a \hbar^2}{G_{\rm N} m^3}}$.  The details of
our extensive DNSs for the GPPE and the SGLPE are given in the Supplemental
Material~\cite{supplement} (see, especially, Table I there).
Figure~\ref{fig:GPPE}(a) shows plots of of $R/R_{\rm Q}$ versus $a/a_{\rm Q}$
for three different values of $G$. In  Fig.~\ref{fig:GPPE}(b) we give  plots of
$M/M_{\rm a}$ versus $R/R_{\rm a}$, from our GPPE DNSs for three different
values of $g$; the trends in these plots are markedly different than those at
$T=0$ (see Fig.(1) in the Supplemental Material~\cite{supplement}). In
Fig.~\ref{fig:GPPE}(c) we plot the scaled energies  $E_{kq}/E_a, \,
E_{int}/E_a, \, E_{\rm G}/E_{\rm a},$ and their total $E/E_{\rm a}$ versus the
scaled temperature $k_{\rm B}T/E_a$ from our SGLPE DNSs; the jumps in these
curves at $k_{\rm B}T/E_{\rm a} \simeq 3.25\times10^{-5}$ suggest a first-order
transition from a collapsed BEC to a tenuous, non-condensed assembly. 

\begin{figure*}[ht]
\centering	
\resizebox{\linewidth}{!}{
\includegraphics[scale=0.5]{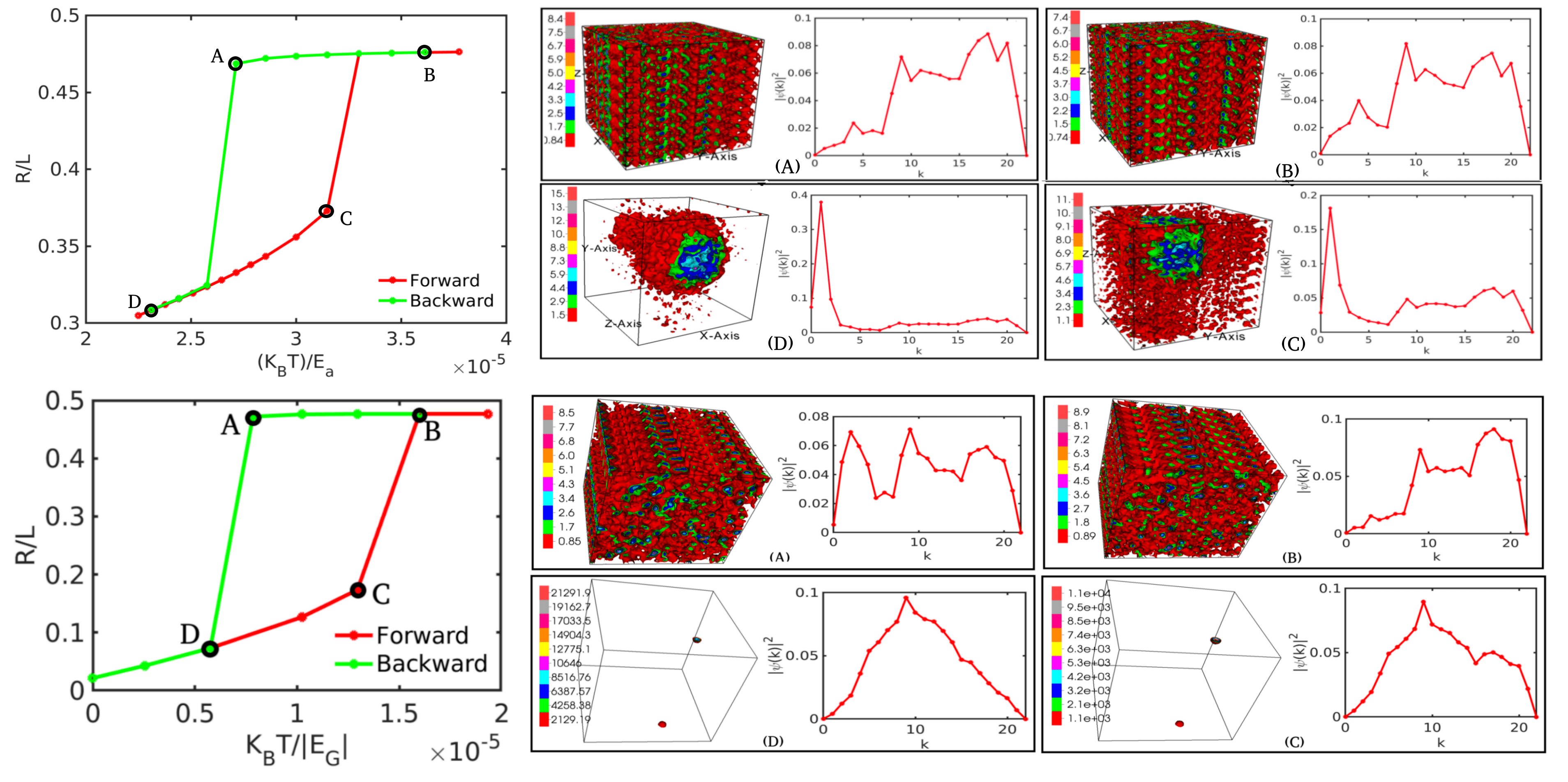}}
\caption{Top left panel: plots of the dimensionless radius ($R/L$) versus 
the dimensionless temperature ($k_{\rm B}T/E_a$), for heating (red) and cooling
(green) runs showing a hysteresis loop. We show ten-level contour plots of
${\left| \psi({\bf x}) \right|}^2$ and the associated spectra
${\left| \psi(k) \right|}^2$ to illustrate, at representative points
on heating and cooling curves in the hysteresis plot, the real-space density
distribution and the $k-$space density spectra
($k_{\rm B}T/E_{\rm a} = 2.7\times 10^{-5}$ and $k_{\rm B}T/E_{\rm a} =
3.62\times10^{-5}$ in panels A and B of top panels, respectively, and
$k_{\rm B}T/E_{\rm a} = 2.3\times {10^{-5}}$ and $k_{\rm B}T/E_{\rm a}=3.16\times10^{-5}$ in panels C and D of bottom panels, respectively). The analogs of these plots, for 
the case $g = 0$, are shown in the panels at the very bottom. In the bottom 
left panel we use  $|E_G|$ at $T=0$ to make the temperature dimensionless. }
\label{fig:rad_hyst}
\end{figure*}

To confirm such a transition, we carry out an SGLPE hysteresis study, whose
results are summarised in the plots in the panels of
Fig.~\ref{fig:rad_hyst}. The left panel of this figure shows  $R/L$ versus
$k_{\rm B}T/E_{\rm a}$; we obtain the red and green curves by, respectively,
increasing and decreasing $k_{\rm B}T/E_{\rm a}$ (often referred to as heating
and cooling runs in statistical mechanics).  In these SGLPE runs, we use the
final steady-state configuration for $\psi({\bf x})$, from the previous
temperature, as the initial condition at the next temperature; clearly, there
is significant hysteresis at the first-order transition from the - collapsed BEC
to the non-collapsed state.  In the right side panels of Fig.~\ref{fig:rad_hyst}, we show ten-level
contour plots of ${\left| \psi({\bf x}) \right|}^2$ and the associated spectra
${\left| \psi(k) \right|}^2$ to illustrate, at representative points on heating
and cooling curves in the hysteresis plot, the real-space density distribution
and the $k-$space density spectra ($k_{\rm B}T/E_{\rm a} = 2.7\times 10^{-5}$
and $k_{\rm B}T/E_{\rm a} = 3.62\times10^{-5}$ in the top panels A and B,
respectively, and $k_{\rm B}T/E_{\rm a} = 2.3\times {10^{-5}}$ and $k_{\rm
B}T/E_{\rm a}=3.16\times10^{-5}$ in the bottom panels C and D,
respectively). From these density distributions and spectra, we conclude that
our system undergoes a first-order transition from a collapsed BEC to a
tenuous, non-condensed assembly. The panel of figures at the very bottom of
Fig.~\ref{fig:rad_hyst} show that this first-order transition occurs at
$g = 0$ too. We expect that this also occurs when $g < 0$, which is the 
appropriate parameter range for axion stars~\cite{Chavanis4,Chavanis5}; the 
$g < 0$ case requires a quintic nonlinearity in Eq.~\ref{eq:GPPE} for stability;
finite-temperature effects can be studied for this case by using the methods that 
we have described above (as we will show in future work).

Do our Fourier-truncated GPPE yield only single collapsed objects? No. We now
show, for the illustrative parameter values $g=20$, $G = 1000$, and $128^3$
collocation points, that this GPPE can also yield long-lived states with
temporal oscillations. In particular, by using an initial condition with two
rotating spherical compact objects, the truncated GPPE dynamics yields a
rotating binary system, which we depict at representative times in
Fig.~\ref{fig:bin_star} and in the Video V6 in the Supplementary
Material~\cite{supplement} (this also gives the initial data).  

Our study goes well beyond earlier studies ~\cite{harkofinite,schivecosmic}
that employ the Thomas-Fermi approximation and include finite-temperature 
effects at the level of 
a non-interacting Bose gas (with $g = 0$). We have shown that the truncated
GPPE can be used effectively to study the
graviational collapse of an assembly of weakly interacting bosons \textit{at 
finite temperature}. Our study of this collapse shows that there is a clear,
thermally
driven \textit{first-order transition} from a non-condensed bosonic gas, at
high $T$, to a condensed BEC phase at low $T$; this transition should be
contrasted with the continuous BEC transition in the weakly interacting Bose
gas, which is described by the GPE, in the absence of gravitation.
Furthermore, we have shown that gravitationally bound, rotating binary objects
can be obtained in our GPPE simulations. Therefore, our work opens up the
possibility of carrying out detailed finite-temperature studies of
self-gravitating bosonic systems, which are potentially relevant for studies of
dark-matter candidates like boson stars and axions.

\begin{figure*}[ht]
\centering	
\resizebox{\linewidth}{!}{
\includegraphics[scale=1.0]{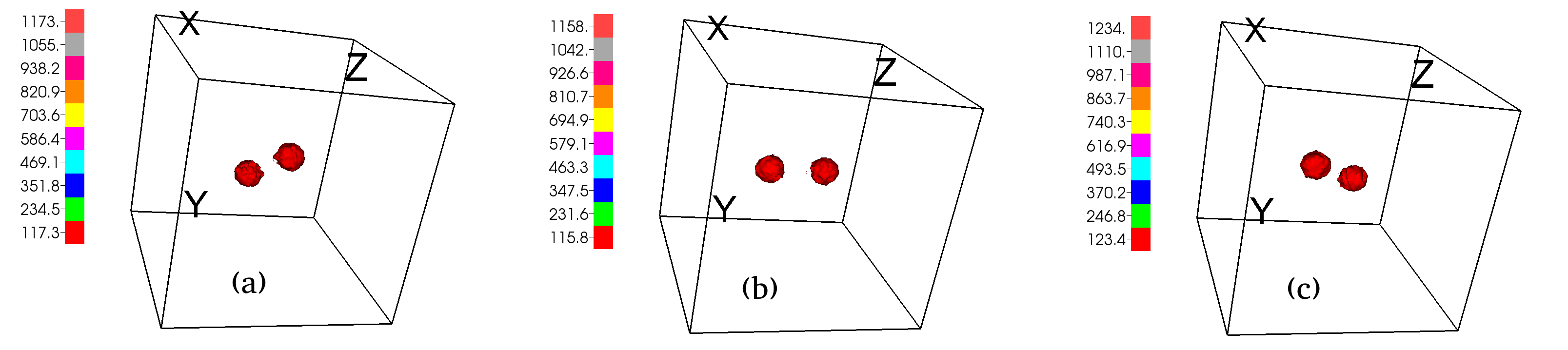}}
\caption{Ten-level contour plots of the $|\psi({\bf x},t)|^2$ (a) at
	$t = 0.018$, (b) at $t = 0.025$, and (c) at $t = 0.03$, for the 
 initial condition for $\psi({\bf x},t)$ given in Eq.(1) of the Supplementary
 Material~\cite{supplement}, showing rotating binary system (see video V6 in
 supplemental material).}\label{fig:bin_star}
\end{figure*}


{\bf Acknowledgements}: We thank DST, CSIR (India), and the Indo French Center 
for Applied Mathematics (IFCAM) for their support.


\begin{thebibliography}{11}
\bibitem{ODell2000} D. O’Dell, S. Giovanazzi, G. Kurizki, and V. M. Akulin, Phys. Rev. Lett. {\bf 84}, 5687 (2000).
\bibitem{RuffiniBonazzola1969} R. Ruffini and S. Bonazzola, Phys. Rev. {\bf 187}, 1767 (1969).
\bibitem{Ingrosso:1991} G. Ingrosso, D. Grasso, and R. Ruffini, Astron. Astro-
phys. {\bf 248}, 481 (1991).
\bibitem{Jetzer1992} P. Jetzer, Physics Reports {\bf 220}, 163 (1992).
\bibitem{Hu2000} W. Hu, R. Barkana, and A. Gruzinov, Phys. Rev. Lett.
{\bf 85}, 1158 (2000).
\bibitem{abdallah2018search} H. Abdallah et al., Phys. Rev. Lett. {\bf 120}, 
201101 (2018).
\bibitem{kachulis2018search} C. Kachulis et al., Phys. Rev. Lett. {\bf 120},
221301v(2018).
\bibitem{apslink} {\url{https://physics.aps.org/articles/v11/48?utm_campaign=weekly&utm_medium=email&utm_source=emailalert}}
\bibitem{lawson2019tunable} M. Lawson et al., Phys. Rev. Lett. {\bf 123}, 
141802 (2019).
\bibitem{Chavanis1} P.-H. Chavanis, Phys. Rev. D {\bf 84}, 043531 (2011).
\bibitem{Chavanis2} P.-H. Chavanis and L. Delfini, Phys. Rev. D {\bf 84},
043532 (2011).
\bibitem{Chavanis3} P.-H. Chavanis, Phy. Rev. D {\bf 94}, 083007 (2016).
\bibitem{Chavanis4} P.-H. Chavanis, Phys. Rev. D {\bf 98}, 023009 (2018).
\bibitem{Chavanis5} P.-H. Chavanis, \textit{Quantum Aspects of Black Holes},
eds. X. Calmet et al, (Springer, 2015), pp. $151–194$.
\bibitem{PhDTutorial} N. P. Proukakis and B. Jackson, Journal of Physics
B: Atomic, Molecular and Optical Physics 41, 203002
(2008).
\bibitem{Berloff14} N. G. Berloff, M. Brachet, and N. P. Proukakis, Proc.
Natl. Acad. Sci. U.S.A. {\bf 111}, 4675 (2014).
\bibitem{Krstulovic11} G. Krstulovic and M. Brachet, Phys. Rev. E {\bf 83},
066311 (2011).
\bibitem{vmrnjp13} V. Shukla, M. Brachet, and R. Pandit, New J. Phys. {\bf 15},
113025 (2013).
\bibitem{supplement} Supplemental Material (2019).
\bibitem{Peebles} P. Peebles, \textit{The large-scale structure of the universe}, 2011 (Princeton University Press, Princeton, NJ, 1980).
\bibitem{falco2013} M. Falco, S. H. Hansen, R. Wojtak, and G. A. Mamon,
Mon. Not. R. Astron. Soc. {\bf 431}, L6 (2013).
\bibitem{KIESSLING2003} M. K.-H. Kiessling, Advances in Applied Mathematics
{\bf 31}, 132 (2003).
\bibitem{Got-Ors} D. Gottlieb, and S. A. Orszag, Numerical Analysis of
Spectral Methods (SIAM, Philadelphia, 1977).
\bibitem{harkofinite} T. Harko, and J.M. Madarassy, J. Cosmol. Astropart. Phys.
{\bf 01} (2012) 025.
\bibitem{schivecosmic} H-Y. Schive, T. Chiueh, and T. Broadhurst, Nat. Phys.
{\bf 10}, 496 (2014)
.
\end{thebibliography}
\end{document}